# NONLINEAR LOCALIZED EXCITATIONS IN HELIX MACROMOLECULES

Natalya Kovaleva, Leonid Manevitch

*Semenov Institute of Chemical Physics, Russian Academy of Science, Moscow 119991, Russia*

*Abstract:* Physico-mechanical properties of polymers in solid state, in particular conditions of their structural transformations, are substantially defined by existence and mobility of elementary nonlinear excitations. The localized oscillatory excitations (breathers) which have been revealed in crystalline PE are important for thermodynamic studies. In the given work the breathers in more complex crystalline of helix macromolecules are analytically investigated for the first time. We show that introduction of adequate variables allows to construct the continuum model describing spatially localized excitations with internal degree of freedom. Characteristic parameters of breathers and conditions of their mobility are defined.

## 1. Introduction

It was shown recently numerically [1] and analytically [2] that vibrational localized excitations can exist in realistic model of zig-zag polymer chain. Such excitations existing in the attenuation bands of linear spectrum, introduce a noticeable contribution into thermal capacity. In this paper we consider essentially more complex case of helix polymer chain (in application to polytetrafluoethylene (PTFE)). The long wavelength localized excitations (acoustic solitons) in helix chains have been studied numerically in the paper [3]. We deal with oscillatory localized excitations which have several deformation components: radial, angular and longitudinal.

## 2. The Model

The ground state of macromolecule PTFE is a helix 1*13/6. In the approximation of "united atoms" we consider the $CF_2$-group as a particle with mass M=50 a.m.u. Then its equilibrium state can be characterized by radius vector: $R = (R_0 \cos(n\Delta\varphi), R_0 \sin(n\Delta\varphi), n\Delta z)$, где $R_0 = 0,41$ Å – radius of helix. The angular and longitudinal steps of the helix are $\Delta\varphi = 12\pi/13$, $\Delta z = 1.298$ Å correspondingly. Let us introduce transversal $r_n$, angular $\varphi_n$ and longitudinal $h_n$ displacements of $n$-th node of the chain with respect to its equilibrium state. Then the Hamilton function of the chain has a view:

$$H = \sum_n \left\{ \frac{1}{2} M[\dot{r}_n^2 + \dot{\varphi}_n^2 (R_0 + r_n)^2 + \dot{h}_n^2] + V(\rho_n) + U(\theta_n) + W(\delta_n) \right\}$$

(1)

Here dot denotes derivative by time $t$, $\rho_n$ – length of $n$-th valence bond, $\theta_n$ – $n$-th valence angle and $\delta_n$ – $n$-th rotation (conformational) angle.

The potential of valence bond is $V(\rho_n) = D_0 [1 - \exp(-\alpha(\rho_n - \rho_0))]$. In accordance with the data [4] $D_0 = 334.72$ kJ/mol, parameter $\alpha = 1.91$ Å.

The potential of valence angle $U(\theta_n) = \frac{K_\theta}{2} [\cos(\theta_n) - \cos(\theta_0)]^2$, where $K_\theta = 529$ kJ/mol [5].

The potential of internal rotation $W(\delta_n)$ characterizes the rotation stiffness of $n$-th valence bond. The PTFE-molecule has 4 isomers. Two of them [trans(+) and trans(−)] have identical energies [ $\delta_1 = \delta_0, \delta_2 = 2\pi - \delta_0, W(\delta_1) = W(\delta_2) = 0$ ], two others [gauche(+) and gauche(-)] possess more high energy [ $\delta_3 \approx 2\pi/3, \delta_4 \approx 4\pi/3, W(\delta_3) = W(\delta_4) > 0$ ]. Potential of internal rotation is characterized by four values: the height of the potential barriers between both trans conformations $\varepsilon_0 = W(0)$, and between trans and gauche conformations $\varepsilon_1 = W(\pi/3)$, the level of the gauche conformation energy $\varepsilon_2 = W(2\pi/3)$, and the height of the barrier between gauche conformations $\varepsilon_3 = W(\pi)$. According to [6] $\varepsilon_0 = 1.674$ kJ/mol, $\varepsilon_1 = 18.42$ kJ/mol, $\varepsilon_2 = 4.186$ kJ/mol and $\varepsilon_3 = 32.02$ kJ/mol.

We approximate the potential of rotation by the expression [3]

$$W(\delta) = [C_1 Z_\beta(\delta) + C_2 Z_\gamma(\delta) - C_3],$$

where $Z_\beta(\delta)$ and $Z_\gamma(\delta)$ are one parametric functions

$$Z_\beta(\delta) = \frac{(1+\beta)\sin^2(\delta/2)}{1-\beta\sin^2(\delta/2)},$$

$$Z_\gamma(\delta) = \left[\frac{(1+\gamma)\sin(3\delta/2)}{1-\gamma\sin(3\delta/2)}\right]^2$$

The magnitudes of the parameters $C_1 = 3.411\,(kJ/mol)^{1/2}$, $C_2 = 2.681\,(kJ/mol)^{1/2}$, $C_3 = 1.294\,(kJ/mol)^{1/2}$, $\beta = 14.6125$, $\gamma = 4.0028\times 10^{-3}$. The potential under given parameters is presented in Fig.1.

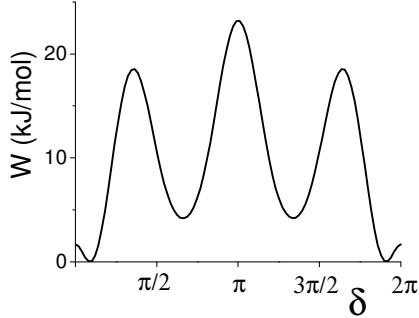

Fig.1. Potential of rotation around the valence bond C-C $W(\delta)$ for macromolecule PTFE.

Dispersion relation for given system is algebraic equation of third order with respect to square of the frequency. Corresponding dispersion curves has three branches: two acoustic $\omega = \omega_t(q), \omega = \omega_l(q)$ and one optic $\omega = \omega_o(q)$ ($\omega_t(q) \le \omega_l(q) \le \omega_o(q)$). These curves are plotted in Fig.2. The lower curve $\omega = \omega_t(q)$ corresponds to dispersion law for torsional phonons, the middle curve $\omega = \omega_l(q)$ – longitudinal phonons and upper one $\omega = \omega_o(q)$ – to high frequency optic phonons [3].

The equations of motion corresponding to Hamilton function (1) can be presented as follows:

$$M\ddot{r}_n - M(R_0 + r_n)\dot{\varphi}_n^2 + \frac{\partial P}{\partial r_n} = 0,$$

$$M(R_0 + r_n)^2 \ddot{\varphi}_n + 2M(R_0 + r_n)\dot{\varphi}_n\dot{r}_n + \frac{\partial P}{\partial \varphi_n} = 0,$$

$$M\ddot{h}_n + \frac{\partial P}{\partial h_n} = 0,$$

where $P = \sum_n \{V(\rho_n) + U(\theta_n) + W(\delta_n)\}$

Let us perform analytical study of nonlinear oscillations in the vicinity of wave number $q = 0$ and $\omega = 520\,cm^{-1}$ (left edge of optic dispersion curve). To extend the normal modes analysis on the nonlinear case let us introduce the slow modulating functions. We use the denotations $r_n = R$, $\varphi_n = \Phi$, $h_n = H$, $r_{n+1} = \tilde{R}$, $\varphi_{n+1} = \tilde{\Phi}$, $h_{n+1} = \tilde{H}$ and consider the power expansions in the vicinity of an arbitrary nth particle using modulating functions $R, \tilde{R}, \Phi, \tilde{\Phi}, H, \tilde{H}$ [2]:

$$r_{n+m} = \cos(mk)\left[R + m\varepsilon\frac{\partial R}{\partial \xi} + \frac{1}{2}m^2\varepsilon^2\frac{\partial^2 R}{\partial \xi^2}\right]$$
$$+ \sin(mk)\left[\tilde{R} + m\varepsilon\frac{\partial \tilde{R}}{\partial \xi} + \frac{1}{2}m^2\varepsilon^2\frac{\partial^2 \tilde{R}}{\partial \xi^2}\right] + \cdots,$$

$$\varphi_{n+m} = \cos(mk)\left[\Phi + m\varepsilon\frac{\partial \Phi}{\partial \xi} + \frac{1}{2}m^2\varepsilon^2\frac{\partial^2 \Phi}{\partial \xi^2}\right]$$
$$+ \sin(mk)\left[\tilde{\Phi} + m\varepsilon\frac{\partial \tilde{\Phi}}{\partial \xi} + \frac{1}{2}m^2\varepsilon^2\frac{\partial^2 \tilde{\Phi}}{\partial \xi^2}\right] + \cdots,$$

$$h_{n+m} = \cos(mk)\left[H + m\varepsilon\frac{\partial H}{\partial \xi} + \frac{1}{2}m^2\varepsilon^2\frac{\partial^2 H}{\partial \xi^2}\right]$$
$$+ \sin(mk)\left[\tilde{H} + m\varepsilon\frac{\partial \tilde{H}}{\partial \xi} + \frac{1}{2}m^2\varepsilon^2\frac{\partial^2 \tilde{H}}{\partial \xi^2}\right] + \cdots,$$

where small parameter $\varepsilon$ characterizes the distance between the particles in the units $\xi = \varepsilon\,z/l_z$, and $z$, $l_z$ - coordinate and helix step along the helix axis, respectively $m = 1, 2, \ldots$.

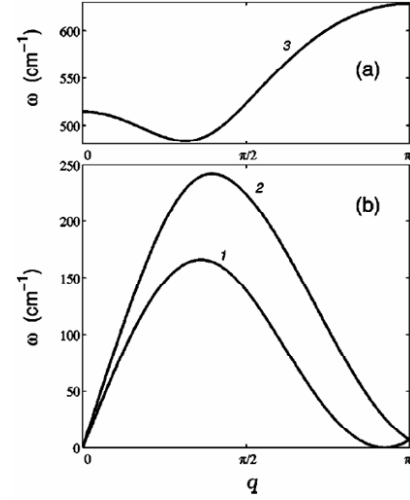

Fig. 2. The frequency spectrum curves $\omega = \omega_t(q)$ (1), $\omega = \omega_l(q)$ (2), and $\omega = \omega_o(q)$ (3), for the spiral isolated macromolecule PTFE.

In the same way we consider the expansions in the vicinity of the $(n+1)$-th particle

$$r_{n+1+m} = \sin(-mk)\left[R + m\varepsilon\frac{\partial R}{\partial \xi} + \frac{1}{2}m^2\varepsilon^2\frac{\partial^2 R}{\partial \xi^2}\right]$$
$$+ \cos(mk)\left[\tilde{R} + m\varepsilon\frac{\partial \tilde{R}}{\partial \xi} + \frac{1}{2}m^2\varepsilon^2\frac{\partial^2 \tilde{R}}{\partial \xi^2}\right] + \cdots,$$

$$\varphi_{n+1+m} = \sin(-mk)\left[\Phi + m\varepsilon\frac{\partial \Phi}{\partial \xi} + \frac{1}{2}m^2\varepsilon^2\frac{\partial^2 \Phi}{\partial \xi^2}\right]$$
$$+ \cos(mk)\left[\tilde{\Phi} + m\varepsilon\frac{\partial \tilde{\Phi}}{\partial \xi} + \frac{1}{2}m^2\varepsilon^2\frac{\partial^2 \tilde{\Phi}}{\partial \xi^2}\right] + \cdots,$$

$$h_{n+1+m} = \sin(-mk)\left[H + m\varepsilon\frac{\partial H}{\partial \xi} + \frac{1}{2}m^2\varepsilon^2\frac{\partial^2 H}{\partial \xi^2}\right]$$
$$+ \cos(mk)\left[\tilde{H} + m\varepsilon\frac{\partial \tilde{H}}{\partial \xi} + \frac{1}{2}m^2\varepsilon^2\frac{\partial^2 \tilde{H}}{\partial \xi^2}\right] + \cdots,$$

The expansions should be considered in the vicinities of two particles because double degeneracy of linear normal modes. By substituting the modulating functions in linearized equations for $r_n, r_{n+1}, \varphi_n, \varphi_{n+1}, h_n, h_{n+1}$ we obtain six partial differential equations taking into account the linear parts of the equations for modulating functions. These equations have to be in accordance with foregoing dispersion relations. In the particular case k=0 the equations are reduced to tree equations:

$$\frac{\partial^2 R}{\partial t^2} + a_1 R + \varepsilon\left[a_4\frac{\partial \Phi}{\partial \xi} + a_5\frac{\partial H}{\partial \xi} - a_7 R^2\right] + \varepsilon^2\left[AA_1\frac{\partial^2 R}{\partial \xi^2} - a_{19}R^3 + R\left(a_{63}\frac{\partial \Phi}{\partial \xi} - a_{51}\frac{\partial H}{\partial \xi}\right)\right] = 0$$

$$\frac{\partial^2 \Phi}{\partial t^2} - \varepsilon b_4\frac{\partial R}{\partial \xi} - \varepsilon^2\left[BB_1\frac{\partial^2 \Phi}{\partial \xi^2} + BB_2\frac{\partial^2 H}{\partial \xi^2} + b_{57}r\frac{\partial R}{\partial \xi}\right] = 0$$

$$\frac{\partial^2 H}{\partial t^2} - \varepsilon c_4\frac{\partial R}{\partial \xi} - \varepsilon^2\left[CC_1\frac{\partial^2 \Phi}{\partial \xi^2} + CC_2\frac{\partial^2 H}{\partial \xi^2} - c_{57}R\frac{\partial R}{\partial \xi}\right] = 0$$

Let us introduce the complex function $u = \frac{\partial R}{\partial \tau} + iR$, where $\tau = \sqrt{a_1}t$. Then the equations of motion are written as

$$\frac{\partial u}{\partial \tau} + iu + \varepsilon\left[A_4\frac{\partial \Phi}{\partial \xi} + A_5\frac{\partial H}{\partial \xi} + \frac{A_7}{4}(u-u^*)^2\right] +$$
$$+ \varepsilon^2\left[-i\frac{A_1}{2}\frac{\partial^2(u-u^*)}{\partial \xi^2} - i\frac{A_{19}}{8}(u-u^*)^3 -\right.$$
$$\left.- i\frac{(u-u^*)}{2}\left(A_{63}\frac{\partial \Phi}{\partial \xi} - A_{51}\frac{\partial H}{\partial \xi}\right)\right] = 0$$

$$\frac{\partial^2 \Phi}{\partial \tau^2} + i\frac{\varepsilon B_4}{2}\frac{\partial(u-u^*)}{\partial \xi} -$$
$$- \varepsilon^2\left[B_1\frac{\partial^2 \Phi}{\partial \xi^2} + B_2\frac{\partial^2 H}{\partial \xi^2} - \frac{B_{57}}{4}(u-u^*)\frac{\partial(u-u^*)}{\partial \xi}\right] = 0$$

$$\frac{\partial^2 H}{\partial \tau^2} + i\frac{\varepsilon C_4}{2}\frac{\partial(u-u^*)}{\partial \xi} - \varepsilon^2\left[C_1\frac{\partial^2 \Phi}{\partial \xi^2} + C_2\frac{\partial^2 H}{\partial \xi^2} + \frac{C_{57}}{4}(u-u^*)\frac{\partial(u-u^*)}{\partial \xi}\right] = 0$$

We introduce, alongside with fast time $\tau_0 = \tau$, the slow times $\tau_1 = \varepsilon\tau, \tau_2 = \varepsilon^2\tau,...$ and expand the unknown functions into power series with respect to small parameter $\varepsilon$:

$u = e^{i\tau_0}(\psi_0 + \varepsilon\psi_1 + \varepsilon^2\psi_2 + ...)$;

$\Phi = (\Phi_0 + \varepsilon\Phi_1 + \varepsilon^2\Phi_2 + ...)$; $H = (H_0 + \varepsilon H_1 + \varepsilon^2 H_2 + ...)$

Selecting the terms of the same order by $\varepsilon$ and equating them to zero one can find:

1) $\varepsilon^0$: $\psi_0 = \psi_0(\tau_1,...,\xi)$, $\Phi_0 = \Phi_0(\tau_1,...,\xi)$, $H_0 = H_0(\tau_1,...,\xi)$

2) $\varepsilon^1$:

$$\frac{\partial \psi_1}{\partial \tau_0} + \frac{\partial \psi_0}{\partial \tau_1} + A_4\frac{\partial \Phi_0}{\partial \xi}e^{-i\tau_0} + A_5\frac{\partial H_0}{\partial \xi}e^{-i\tau_0}$$
$$+ \frac{A_7}{4}e^{i\tau_0}\psi_0^2 - \frac{A_7}{2}e^{-i\tau_0}\psi_0\psi_0^* + \frac{A_7}{4}e^{-3i\tau_0}\psi_0^{*2} = 0$$
(2)

$$\frac{\partial^2 \Phi_1}{\partial \tau_0^2} + 2\frac{\partial^2 \Phi_0}{\partial \tau_1 \partial \tau_0} + i\frac{B_4}{2}\left[e^{i\tau_0}\frac{\partial \psi_0}{\partial \xi} - e^{-i\tau_0}\frac{\partial \psi_0^*}{\partial \xi}\right] = 0,$$
(3)

$$\frac{\partial^2 H_1}{\partial \tau_0^2} + 2\frac{\partial^2 H_0}{\partial \tau_1 \partial \tau_0} + i\frac{C_4}{2}\left[e^{i\tau_0}\frac{\partial \psi_0}{\partial \xi} - e^{-i\tau_0}\frac{\partial \psi_0^*}{\partial \xi}\right] = 0.$$
(4)

To avoid the appearance of growing (secular) terms, we suppose:

$\frac{\partial \psi_0}{\partial \tau_1} = 0$, $\psi_0 = \psi_0(\tau_2,...,\xi)$, $\Phi_0 = \Phi_0(\tau_2,...,\xi)$, $H_0 = H_0(\tau_2,...,\xi)$

Integrating these equations (2-3) one can obtain

$$\psi_1 = -ie^{-i\tau_0}\left[A_4\frac{\partial \Phi_0}{\partial \xi} + A_5\frac{\partial H_0}{\partial \xi}\right] +$$
$$+ i\frac{A_7}{4}\left[\psi_0^2 e^{i\tau_0} + 2\psi_0\psi_0^* e^{-i\tau_0} - \frac{1}{3}\psi_0^{*2}e^{-3i\tau_0}\right]$$

$$\Phi_1 = i\frac{B_4}{2}\left[\frac{\partial \psi_0}{\partial \xi}e^{i\tau_0} - \frac{\partial \psi_0^*}{\partial \xi}e^{-i\tau_0}\right]$$

$$H_1 = i\frac{C_4}{2}\left[\frac{\partial \psi_0}{\partial \xi}e^{i\tau_0} - \frac{\partial \psi_0^*}{\partial \xi}e^{-i\tau_0}\right]$$

Substituting the expressions for $\psi_1, \Phi_1, H_1$ into equations of order $\varepsilon^2$ selecting the secular terms and equating them to zero, we come to the main asymptotic approximation, that are the conditions of absence of growing(secular) terms in the solution.

For $\varepsilon^2$ ::

$$\frac{\partial \psi_0}{\partial \tau_2} - iP_1\frac{\partial^2 \psi_0}{\partial \xi^2} + iP_2|\psi_0|^2\psi_0 + iP_3\frac{\partial H_0}{\partial \xi}\psi_0 - iP_4\frac{\partial \Phi_0}{\partial \xi}\psi_0 = 0 \quad (5)$$

$$P_5\frac{\partial^2 H_0}{\partial \xi^2} - P_6\frac{\partial^2 \Phi_0}{\partial \xi^2} - P_7\frac{\partial |\psi_0|^2}{\partial \xi} = 0 \quad (6)$$

$$P_8\frac{\partial^2 H_0}{\partial \xi^2} - P_9\frac{\partial^2 \Phi_0}{\partial \xi^2} + P_{10}\frac{\partial |\psi_0|^2}{\partial \xi} = 0 \quad (7)$$

where $P_1 = \frac{1}{2}(A_1 - A_4B_4 - A_5C_4) = 0.124$;

$P_2 = \frac{1}{24}(9A_{19} + 10A_7^2) = 244.7$; $P_3 = \frac{1}{2}(A_{51} - 2A_5A_7) = 14.345$;

$P_4 = \frac{1}{2}(A_{63} + 2A_4A_7) = 0.1036$; $P_5 = B_4A_5 - B_2 = 0.0035$;

$P_6 = B_1 - A_4B_4 = 0.002$;

$P_7 = \frac{(2A_7B_4 + B_{57})}{4} = 0.308$;

$P_8 = C_4A_5 - C_2 = 0.105$; $P_9 = C_1 - A_4C_4 = 0.001$;

$P_{10} = \frac{(C_{57} - 2A_7C_4)}{4} = 7.17$.

Then we express $\frac{\partial H_0}{\partial \xi}, \frac{\partial \Phi_0}{\partial \xi}$ via $\psi_0$ using equations (6) and (7):

$$\frac{\partial \Phi_0}{\partial \xi} = Q_1|\psi_0|^2, \quad (8)$$

$$\frac{\partial H_0}{\partial \xi} = Q_2|\psi_0|^2, \quad (9)$$

where $Q_1 = \frac{P_7P_8 + P_{10}P_5}{P_5P_9 - P_8P_6}$, $Q_2 = \frac{P_7P_9 + P_{10}P_6}{P_5P_9 - P_8P_6}$

Substitution of (8), (9) into (5) leads to final asymptotic equation:

$$\frac{\partial \psi_0}{\partial \tau_2} - iP_1\frac{\partial^2 \psi_0}{\partial \xi^2} - iQ|\psi_0|^2\psi_0 = 0, \quad (10)$$

which is nonlinear Schrodinger equation – one of the fundamental equations of Nonlinear Dynamics.

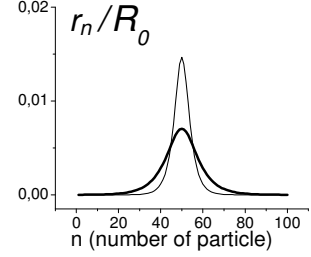

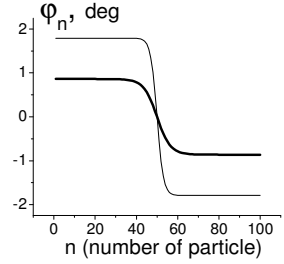

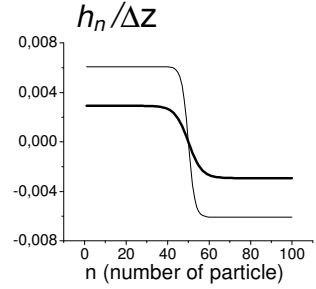

Fig.3. Profiles for radial, angle and longitudinal variables. Thin line $\Delta\omega = -1.3$, thick line $\Delta\omega = -0.3$; $U = 0$, $\varepsilon = 0.1$.

The parameters are: $P_1 = \frac{1}{2}(A_1 - A_4B_4 - A_5C_4) = 0.124$;

$$Q = P_4\frac{P_7P_8 + P_{10}P_5}{P_9P_5 - P_6P_8} - P_2 - P_3\frac{P_7P_9 + P_{10}P_6}{P_5P_9 - P_8P_6} = 750$$

After change $\psi_0 = e^{ik_r\xi - i\Delta\omega\tau_2}v(\xi - U\tau_2)$, if $U = 2P_1k_r$ we obtain that equation (10) is reduced to nonlinear ordinary equation:

$v_1'' - \alpha v_1 + v v_1^3 = 0$, where $\alpha = k_r^2 - \frac{\Delta\omega}{P_1}$, $v = \frac{Q}{P_1}$

Its spatially localized solution has a view

$v = (\frac{2\alpha}{v})^{1/2}\text{sech}(\alpha^{1/2}(\xi - U\tau_2))$

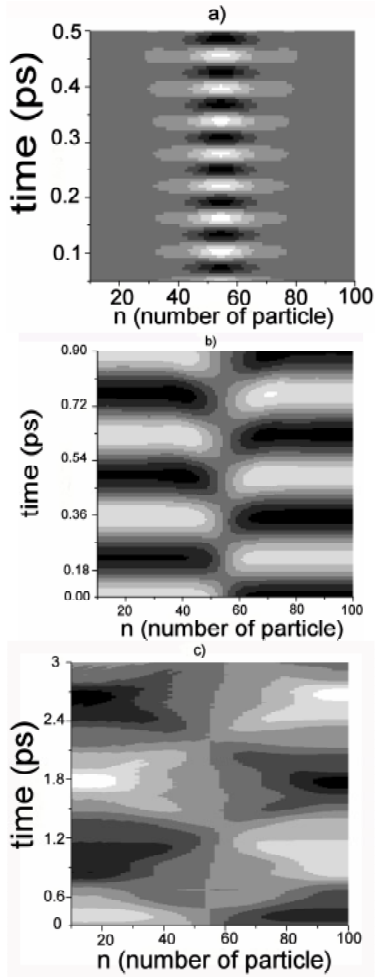

Fig.4. Time-coordinate maps for radial $r_n$ (a), angle $\varphi_n$ (b) and longitudinal $h_n$ (c) variables.

if $\alpha > 0$, $\nu > 0$. To satisfy these conditions it is necessary to provide $\Delta\omega < k_r^2 P_1 = \dfrac{U^2}{4P_1}$

Returning to initial variables, one can obtain:

$r = \varepsilon (\dfrac{2\alpha}{\nu})^{1/2} \operatorname{sech}(\alpha^{1/2}(\xi - U\varepsilon^2 \sqrt{a_1} t)) \sin(\sqrt{a_1} t(1 - \Delta\omega\varepsilon^2) + k_r\xi)$

$\varphi = \dfrac{2\varepsilon Q_1 \sqrt{\alpha}}{\nu} \operatorname{th}(\alpha^{1/2}(\xi - U\varepsilon^2 \sqrt{a_1} t))$

$h = \dfrac{2\varepsilon Q_2 \sqrt{\alpha}}{\nu} \operatorname{th}(\alpha^{1/2}(\xi - U\varepsilon^2 \sqrt{a_1} t))$

The plots describing the spatial profiles of radial angular and longitudinal coordinates corresponding to localized excitation are presented in Fig.3.

It is interesting to note that breathers-like behavior of radial coordinate is accompanied by kink-like spatial dependence of angular and longitudinal coordinates. Corresponding time-coordinate maps are plotted in Fig.4.

## 3. Conclusion

The localized vibration excitation in helix polymer chain has been found analytically for the first time. For this the asymptotic approach with using the complex presentation of equations of motion and multiple scale expansions has been applied. At that, we deal with of nonlinear modulation of linear vibrations possessing optic branch of dispersion curve .It is shown that this excitation which is a bright breather has a rather complex structure corresponding to a combination of three types of deformation: radial, longitudinal and rotational. These results were confirmed by computer simulation with using of initial discrete equations of motion.

## 4 Acknowledgements

This research was supported by CRDF (grant RUB2-2920-MO-07).